\documentclass[final,3p,times,twocolumn]{elsarticle}
\usepackage{slashed}
\usepackage{multirow}
\usepackage{url}
\usepackage{hyperref}
\usepackage[normalem]{ulem}
\usepackage{xcolor}
\usepackage{amsmath}

\makeatletter
\def\ps@pprintTitle{%
 \let\@oddhead\@empty
 \let\@evenhead\@empty
 \def\@oddfoot{}%
 \let\@evenfoot\@oddfoot}
\makeatother

\begin{document}

\begin{frontmatter}

\title{Updated LHC bounds on MUED after Run 2\\}


\author[label1]{Marvin M. Flores}
\author[label2]{Jong Soo Kim}
\author[label4]{Krzysztof Rolbiecki}
\author[label5]{Roberto Ruiz de Austri Bazan}

\address[label1]{National Institute of Physics, University of the Philippines, Diliman, Quezon City, Philippines}
\address[label2]{National Institute for Theoretical Physics and School of Physics, University of the Witwatersrand, Johannesburg, South Africa}
\address[label4]{Faculty of Physics, University of Warsaw, Warsaw, Poland}
\address[label5]{Instituto de Física Corpuscular, IFIC-UV/CSIC, 46980 Paterna, Valencia, Spain}

\begin{abstract}
{We present updated LHC limits on the minimal universal extra dimensions (MUED) model from the Run 2 searches. We scan the parameter space against a number of searches implemented in the public code \textsc{CheckMATE} and derive up-to-date limits on the MUED parameter space from 13 TeV searches. The strongest constraints come from a search dedicated to squarks and gluinos with one isolated lepton, jets and missing transverse energy. In the procedure we take into account initial state radiation and stress its importance in the MUED searches, which is not always appreciated. }
\end{abstract}

\end{frontmatter}

\section{Introduction}

Over the years, the LHC collaborations have put emphasis on probing various incarnations of low energy supersymmetry (SUSY) such as the minimal supergravity (mSUGRA)~\cite{Chamseddine:1982jx}, gauge-mediated SUSY \cite{gaugemediated}, anomaly-mediated SUSY \cite{anomaly1,anomaly2}, phenomenological Minimal Supersymmetric Standard Model (pMSSM) \cite{Conley:2010du}, and the electroweak-MSSM \cite{ATLAS:2016dei}, usually represented by simplified models. After extensively collecting data for almost a decade, however, no convincing signal of supersymmetric models has been seen. This naturally led to a growing interest in various alternative beyond-the-Standard-Model (BSM) scenarios. Among these are the Universal Extra Dimensions (UED) \cite{Appelquist:2000nn,Arkani-Hamed:1998jmv,Antoniadis:1998ig,Randall:1999ee,Giudice:1998ck,Han:1998sg,Kakuda:2013kba} which represent a simple extension of the Standard Model (SM) that include a dark matter candidate and are also testable at the LHC. Here we focus on the minimal version, Minimal UED (MUED) \cite{Cembranos:2007vg}. The phenomenology of MUED model bears some similarities to SUSY in that each particle of the SM has a heavier Kaluza-Klein (KK) partner with the same gauge quantum numbers and couplings. However, in contrast to SUSY, KK-particles have the same spin as their SM counterparts. The lightest particle in the KK spectrum is assumed stable and can be a dark matter candidate.

Consequently, the usual collider signatures of the MUED are similar to the SUSY missing energy signatures with additional jets and leptons. The most important differences are the degenerate spectrum of a typical MUED model and several times higher cross-section, for the same overall mass scale. The definitive differentiation between these models requires a detailed analysis of angular and invariant mass distributions of the final state products~\cite{Barr:2004ze,Athanasiou:2006ef,Smillie:2005ar,Wang:2006hk,Burns:2008cp,MoortgatPick:2011ix}. Thus, SUSY searches should provide a high discovery and exclusion potential for the MUED model. Since ATLAS and CMS have performed many searches targeting SUSY, these can be recast to provide bounds on the MUED model as was done in \cite{ATLAS:2015gky,jongpaper,indianpaper1,indianpaper2}. In addition, non-SUSY-like searches can provide bounds as well, for example the CMS search for a di-lepton resonance at $\sqrt{s} = 8$ TeV with 20.6 fb$^{-1}$ which can be used to set limits on the second KK-photon \cite{dilepton}.

A huge advantage of MUED is its predictivity and simplicity since it only depends on two parameters, the compactification radius of the extra dimension $R$, and a cutoff scale $\Lambda$ at which it is replaced by a high-energy theory. The inverse compactification radius $R^{-1}$ sets the mass scale of the first KK excitations while $\Lambda\cdot R$ controls the allowed KK modes present in the spectrum below the cutoff. Existing bounds prior to this paper exclude models with $R^{-1}$ values of up to 1500 GeV \cite{jongpaper,indianpaper2}.

As already mentioned, MUED spectra tend to have smaller mass splittings than supersymmetric spectra which results in softer final state products and thus decreased missing transverse energy. As a result, exclusion limits rely on initial state radiation (ISR) which provides transverse boost to the hard scattering system, especially for low values of $\Lambda\cdot R$. As pointed out in Refs.~\cite{jamie2,Drees:2012dd} this may result in significant QCD uncertainties when setting limits. Therefore, in this Letter we also explore how different settings in Monte-Carlo (MC) event generator \textsc{Pythia 8}~\cite{pythia}, affect constraints and their uncertainties when setting limits on the MUED parameter space. We note that this question was not considered in the earlier studies~\cite{jongpaper,indianpaper1}. We do not compare different MC event generators (e.g. \textsc{Herwig}~\cite{Bellm:2015jjp} or \textsc{Sherpa} \cite{Hoche:2014kca}) since after the matching procedure, the differences between the different parton shower event generators should be marginal \cite{Bornhauser:2009ru}. We rather focus on the change of the parton shower scale uncertainty in Pythia as discussed in \cite{Dreiner:2012gx,Dreiner:2012sh}. Additionally, we take advantage of several ATLAS studies using full luminosity collected in Run 2.

This paper is arranged as follows: Section \ref{S:2} gives a brief overview of the MUED model. Section \ref{S:3} then explains the technical details of ISR in MC simulation and the matching procedure. Section \ref{S:4} provides details of the numerical simulations and analysis as well as a discussion of our bounds in the $R^{-1}$--$\Lambda \cdot R$ plane.

\section{MUED Overview}
\label{S:2}

We focus on the LHC phenomenology of the first KK level excitation. In that case, each chiral SM fermion ($Q_i$, $u_i$, $d_i$, $L_i$, $e_i$) (where $i$ is the SM generation index) has one Dirac-fermion partner ($Q_i^{(1)}$, $u_i^{(1)}$, $d_i^{(1)}$, $L_i^{(1)}$, $e_i^{(1)}$), each SM gauge boson $g_\mu$, $W_\mu$, $B_\mu$ has a massive boson partner $g_\mu^{(1)}$, $W_\mu^{(1)}$, $B_\mu^{(1)}$, and the Higgs partner sector contains a scalar, a pseudo-scalar and a charged partner ($h^{(1)}$, $A_0^{(1)}$, $H_\pm^{(1)})$. The couplings between new states are equal to the SM couplings of their partners. The masses of the KK modes are at the tree level given by \cite{Murayama:2011hj,Belyaev:2012ai}
\begin{align}
\label{eq:mass1}
&m_n = \sqrt{(n/R)^2 + m_\mathrm{SM}^2} ~~~n \geq 1~~\mbox{(for bosons)},\\
&m_n =~n/R + m_\mathrm{SM} ~~~n \geq 1~~\mbox{(for fermions)},\label{eq:mass2}
\end{align}
where $n$ is the KK level, $R$ is the compactification radius and $m_\mathrm{SM}$ the mass of SM state. Since the KK modes at $n=1$ are the lightest, they can be abundantly produced at the LHC and will be our main focus. Equations~(\ref{eq:mass1}) and \eqref{eq:mass2} suggests a very compressed mass spectrum. However, at the loop level, the near-mass-degeneracy is partially lifted with splittings growing with $\Lambda\cdot R$, making the KK-gluon the heaviest state and the KK-partner of the $U(1)_B$ gauge boson the lightest state at each KK level.

We restrict ourselves to the strong production of the colored KK-modes, i.e. KK-gluons and KK-quarks,
\begin{equation}
pp\rightarrow g^{(1)}g^{(1)},~pp\rightarrow \mathcal{Q}_i^{(1)}\mathcal{Q}_j^{(1)},~pp\rightarrow g^{(1)}\mathcal{Q}_j^{(1)},
\end{equation}
where $\mathcal{Q}^{(1)} = Q^{(1)}, q^{(1)}$ and $Q^{(1)}$ denote the SU(2) doublet quark partners (or their anti-particles) and $q^{(1)} = u^{(1)}, d^{(1)}$ are the SU(2) singlet quark partners (or their anti-particles). The production of other states is suppressed by the electroweak couplings.\footnote{Although this is not the case if we consider the second KK level \cite{dilepton}.}

\begin{figure}[t]
\centering
\includegraphics[width=1.0\linewidth]{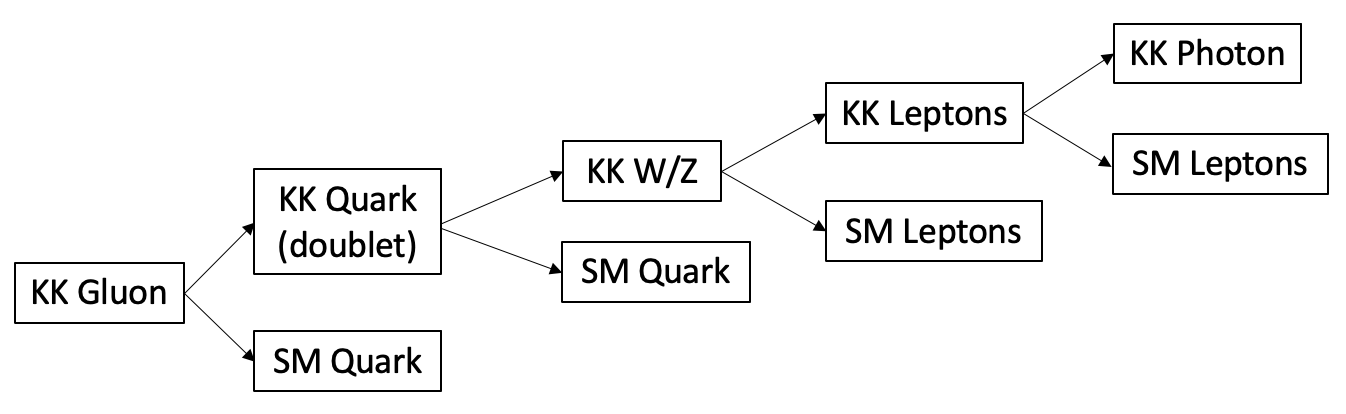}
\caption{A typical decay chain of the KK particles following their dominant branching ratios.}
\label{fig:decay}
\end{figure}

The typical decay patterns of the KK-states, shown diagrammatically in Fig.~\ref{fig:decay}, proceed as follows. The KK-gluon is the heaviest particle and decays into KK-doublet and singlet quarks with about equal branching ratio. The KK-quark decay modes depend mainly on its SU(2) charge. The SU(2) singlet KK-quark directly decays to the KK-photon which is stable and the lightest KK particle (LKP). The SU(2) doublet KK-quarks mainly decay into the KK-gauge-bosons $W_\pm^{(1)}$ and $Z^{(1)}$. The KK-$W$ and $Z$ bosons are lighter than the KK-quarks and so decay mostly into leptonic KK-states which in turn decay further to the LKP and a lepton. Hence, events usually have a relatively large lepton multiplicity, multiple jets, and missing transverse momentum ($\slashed{E}_T$) in the final states, although all decay products can be relatively soft due to the compressed spectrum. Explicit mass spectra are shown in Table~\ref{tab:xs} for various benchmark points with $\Lambda\cdot R = 5$ together with cross sections at $\sqrt{s} = 13$ TeV.

\begin{table*}[t]
\centering
\caption{The total production cross sections, $\sqrt{s}=13$ TeV, and mass spectra for various $R^{-1}$ with $\Lambda\cdot R = 5$.}
\begin{tabular}{|c|c|c|c|c|c|c|c|}
\hline
             &                  & \multicolumn{6}{c|}{Mass   {[}GeV{]}}  \\ \hline
$R^{-1}$ {[}GeV{]} & Total $\sigma$ {[}fb{]} & $g^{(1)}$ & $Q^{(1)}$ & $q^{(1)}$ & $W^{(1)}$/$Z^{(1)}$ & $L^{(1)}$ & $\gamma^{(1)}$\\ \hline
1700         & 76.858           & 1970.48  & 1878.22      & 1854.21      & 1755.49 & 1727.33 & 1698.55\\ \hline
1800         & 45.614           & 2086.39  & 1988.7       & 1963.28      & 1858.54 & 1828.94 & 1798.40\\ \hline
1900         & 27.209           & 2202.3   & 2099.19      & 2072.35      & 1961.60 & 1930.54 & 1898.26\\ \hline
2000         & 16.335           & 2318.22  & 2209.67      & 2181.14      & 2064.68 & 2032.15 & 1998.11\\ \hline
\end{tabular}
\label{tab:xs}
\end{table*}

\section{ISR and the matching procedure}
\label{S:3}

For models with compressed mass spectra, a large recoil is necessary to produce a signature with relatively low SM background. The recoil provides large missing transverse momentum (if there are heavy particles escaping detection) and gives additional transverse boost to leptons if present. The recoil is provided by the initial state radiation from incoming partons. At the MC level, the ISR can be produced in the hard matrix elements or in parton shower. However, modeling ISR using a parton shower approach results in large uncertainties which are mainly due to a choice of a starting scale for parton shower evolution. This is known to have a large effect in squark and gluino production, where the uncertainty can be illustrated by varying between the ''wimpy" and ''power" shower settings, which respectively refers to the 
\textsc{SpaceShower:pTmaxMatch = 1} and 
\textsc{SpaceShower:pTmaxMatch = 2} settings in \textsc{Pythia 8} \cite{jamie2,jamie1,skands}.

In \textsc{Pythia 8} for the ``default'' shower setting (which is 
\textsc{SpaceShower:pTmaxMatch = 0}) $pT_{max}$ is chosen to be the factorization scale for internal processes and the scale value for Les Houches input if the final state of the hard process (not counting subsequent resonance decays) contains at least one quark (excluding top), gluon or photon. If this is not the case, the emissions are allowed to go all the way up to the kinematical limit. This is because in the former set of processes the ISR emission of yet another quark, gluon or photon could lead to double-counting, while no such danger exists in the latter case. The ``wimpy'' setting, on the other hand, always uses the factorization scale for an internal process and the scale value for Les Houches input, i.e.\ the lower value. This avoids double-counting, but may leave out some emissions that ought to have been simulated. Finally, the ``power'' setting always allows emissions up to the kinematical limit which will simulate all possible event topologies, but may lead to double-counting \cite{pythia}.

\begin{figure}
\centering
Leading Jet
\includegraphics[width=1.0\linewidth]{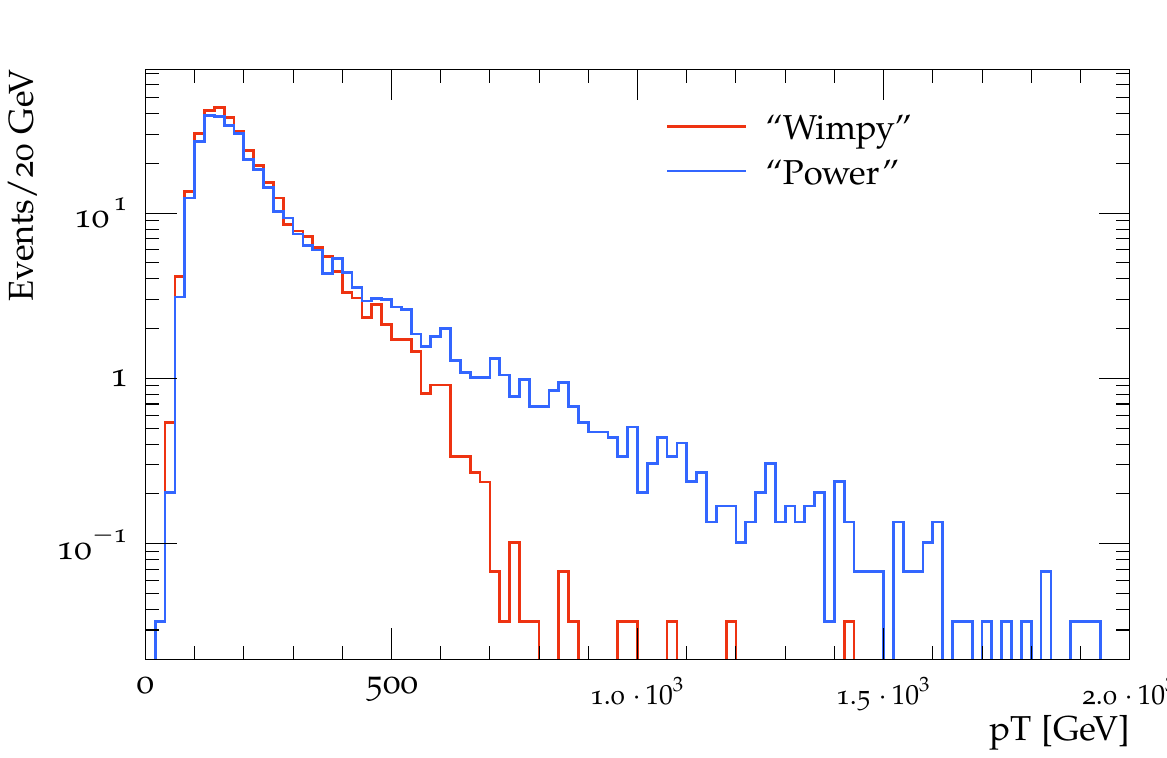}
\caption{A comparison of the leading jet $p_T$ distribution between the ``wimpy" and ``power" settings for the production of KK gluons using \textsc{Pythia 8} for the benchmark point $R^{-1} = 1700$ GeV and $\Lambda\cdot R = 5$ with 10,000 events at 13 TeV normalised to an integrated luminosity of 150 fb$^{-1}$.}
\label{fig:uncertainty}
\end{figure}

Since MUED has a compressed spectrum, especially in the low $\Lambda \cdot R$ region of its parameter space as discussed in Section \ref{S:2} and suggested by Eqs.~(\ref{eq:mass1}) and \eqref{eq:mass2}, we must rely on the ISR and therefore expect the variation in model limits when using different shower settings. This can be seen in Fig. \ref{fig:uncertainty} where the production of KK-gluon was simulated using the \textsc{Pythia 8} setup found in the Appendix C of Ref.~\cite{indianpaper2} for the benchmark point $R^{-1} = 1700$ GeV and $\Lambda\cdot R = 5$ with 10,000 events at $\sqrt{s}=13$ TeV and analysed using \textsc{Rivet} \cite{rivet}.

Therefore, if our search strategy relies on the presence of ISR, we cannot solely use parton shower simulation for determination of limits. The simulation of additional emissions at the hard matrix element level matched to the parton shower is the method that reduces the uncertainty due to ISR modelling.

\section{Numerical Analysis}
\label{S:4}

The samples, either with or without matching, of QCD production of KK-particles, as outlined in Section~\ref{S:2}, are generated using \textsc{MadGraph 5} with the UFO~\cite{Alloul:2013bka} implementation of MUED~\cite{Cheng:2002ab,Datta:2010us,Cheng:2002iz,mued-ufo} and the parton distribution function (PDF) set \textsc{NN23LO1} \cite{Ball:2013hta,Buckley:2014ana}. For the matched samples we use MLM matching \cite{mlm2} implemented in \textsc{Pythia 8} and matrix element with up to one extra parton. Following the guidelines from \textsc{MadGraph} for MLM matching, we checked that varying \textsc{xqcut} and \textsc{qcut} neither change the matched cross sections nor any jet distributions. Moreover, we checked that the matched cross section is consistent with the unmatched cross section. The matching scale was also varied in the range between $(m_1 + m_2)/12$ and $(m_1 + m_2)/3$,  where $m_1$ and $m_2$ are the masses of the final state KK-particles and the resulting total cross section was stable regardless of the parton shower setting. In the following we present the results for a matching scale equal to $(m_1 + m_2)/8$. With these settings, we scanned the parameter space $R^{-1}$--$\Lambda\cdot R$ producing 50,000 events at each grid point.

The events where then fed into \textsc{CheckMATE} \cite{checkmate,Kim:2015wza,Drees:2013wra} for detector simulation and checked for exclusion against 29 ATLAS and CMS searches (the full list is given in the Appendix), corresponding to version \textsc{2.0.30}. Each analysis typically contains a large number of signal regions which target different mass hierarchies and final state multiplicities. For a given search, the \textit{best signal region} for each point in the parameter space is defined by \textsc{CheckMATE} as the one with the largest \textit{expected} exclusion potential. This criterion is then repeated to select the \textit{best search} which is defined as the search whose \textit{best signal region} provides the strongest exclusion. This means that the best \textit{observed} limit is not always used but it ensures that the result is less sensitive to downward fluctuations in the data that are bound to be present when scanning over many searches and many signal regions. Once the \textit{best search} is found for a given point in the parameter space, the signal yield in our model is then compared to the observed limit at 95\% confidence level (CL), 
\begin{equation}
\label{eq:r}
r = \frac{S - 1.64\cdot\Delta S}{S^{95}_{\mbox{obs}}} ,
\end{equation}
where $S$ denotes the number of signal events, $\Delta S$ is the 1-$\sigma$ uncertainty on $S$ combining the statistical MC uncertainty and the 8\% systematic cross section uncertainty\footnote{This approximately amounts to a size of the uncertainty for the next-to-leading order corrections to the cross section~\cite{Freitas:2017hov}.} and $S^{95}_{\mbox{obs}}$ is the observed 95\% CL exclusion limit. The quantity $S - 1.64\cdot\Delta S$ corresponds to the 95\% CL lower bound on our prediction for the number of signal events, which ensures that the limits we set are conservative. The $r$ value is only calculated for the expected best signal region. \textsc{CheckMATE} does not, by default, combine signal regions nor analyses in order to optimize exclusion since the correlations between different searches and signal regions are not known in most cases. We consider a model point as excluded if $r > 1$.

The searches which turn out to be the most sensitive to the MUED model are: \texttt{atlas\_conf\_2019\_040} (2-6 jets and $\slashed{E}_T$, $\mathcal{L} = 139$ fb$^{-1}$) \cite{atlas040}, \texttt{atlas\_2101\_01629} (1 lepton, jets and $\slashed{E}_T$, $\mathcal{L} = 139$ fb$^{-1}$)~\cite{atlas2101}, and \texttt{cms\_sus\_16\_039} (2 same-sign leptons or at least 3 leptons plus $\slashed{E}_T$, $\mathcal{L} = 35.9$ fb$^{-1}$) \cite{cms039}. These analyses are also summarized in Table~\ref{tab:analysis}.

\begin{table}[t]
\caption{Relevant $\sqrt{s} = 13$ TeV analyses used in our study. The middle column denotes the target final state while the last column shows the total integrated luminosity.}
\begin{tabular}{|l|l|l|}
\hline
Analysis               & Final State  & $\mathcal{L}$ {[}fb$^{-1}${]} \\ \hline
\texttt{atlas\_2101\_01629} \cite{atlas2101}      & 1$l$ + jets + $\slashed{E}_T$   & 139                        \\
\texttt{atlas\_conf\_2019\_040} \cite{atlas040} & jets + $\slashed{E}_T$   & 139                         \\
\texttt{cms\_sus\_16\_039} \cite{cms039}      & leptons + $\slashed{E}_T$ & 35.9                        \\ \hline
\end{tabular}
\label{tab:analysis}
\end{table}

\begin{figure}[t]
\centering\includegraphics[width=1.0\linewidth]{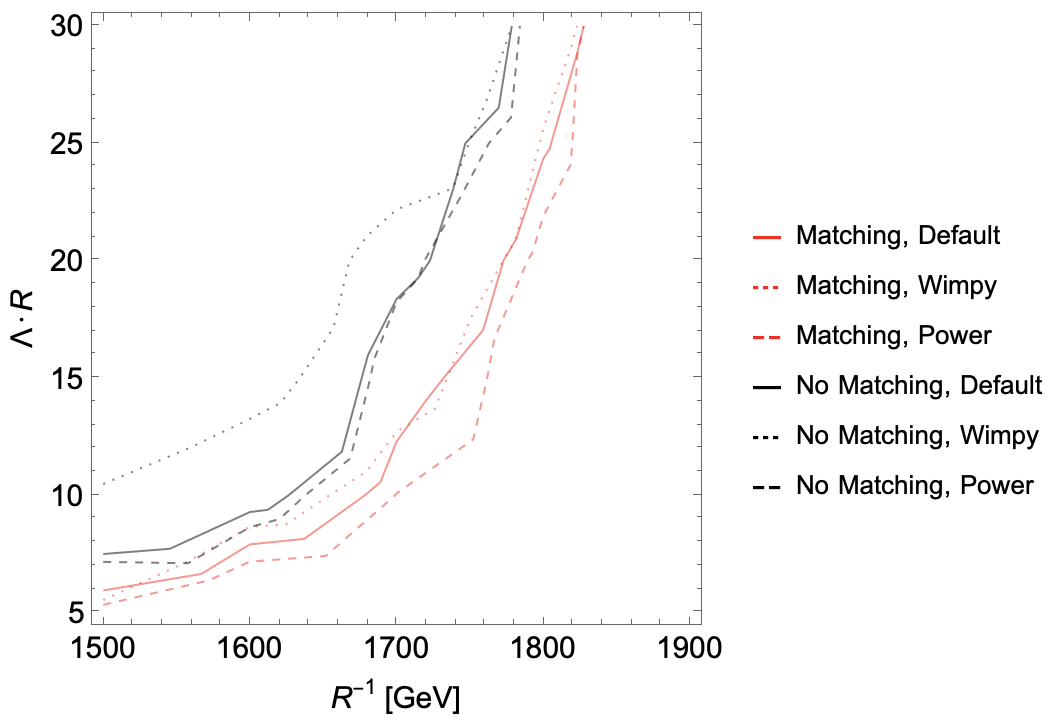}
\caption{Limit comparison of shower-only and matched samples using the ``wimpy" and ``power" mode for the signal region \texttt{4J\_lx\_bveto\_1600} from the ATLAS search \texttt{atlas\_2101\_01629}~\cite{atlas2101}. The difference is noticeable for the unmatched case at lower values of $\Lambda\cdot R$ which corresponds to a compressed mass spectrum.}
\label{fig:matching}
\end{figure}

We start our numerical analysis by comparing an exclusion limit in the $R^{-1}$--$\Lambda\cdot R$ plane obtained for different showering settings in both matched and unmatched samples as discussed in Section~\ref{S:3}. Figure~\ref{fig:matching} shows limit comparison of shower-only and matched samples using the ``wimpy" and ``power" mode for the signal region \texttt{4J\_lx\_bveto\_1600} from the ATLAS search \texttt{atlas\_2101\_01629}~\cite{atlas2101}. As can be seen the limit derived from the matched samples is stronger regardless of the shower setting. A band between power and wimpy setting can be understood as an uncertainty in the exclusion limit due to showering. As expected from the discussion in Section~\ref{S:3} it becomes more pronounced towards lower values of of  $\Lambda\cdot R$ which correspond to more compressed mass spectra. For the unmatched case, the uncertainty band widens significantly in the low $\Lambda\cdot R$.

\begin{figure}[t]
\centering\includegraphics[width=1.0\linewidth]{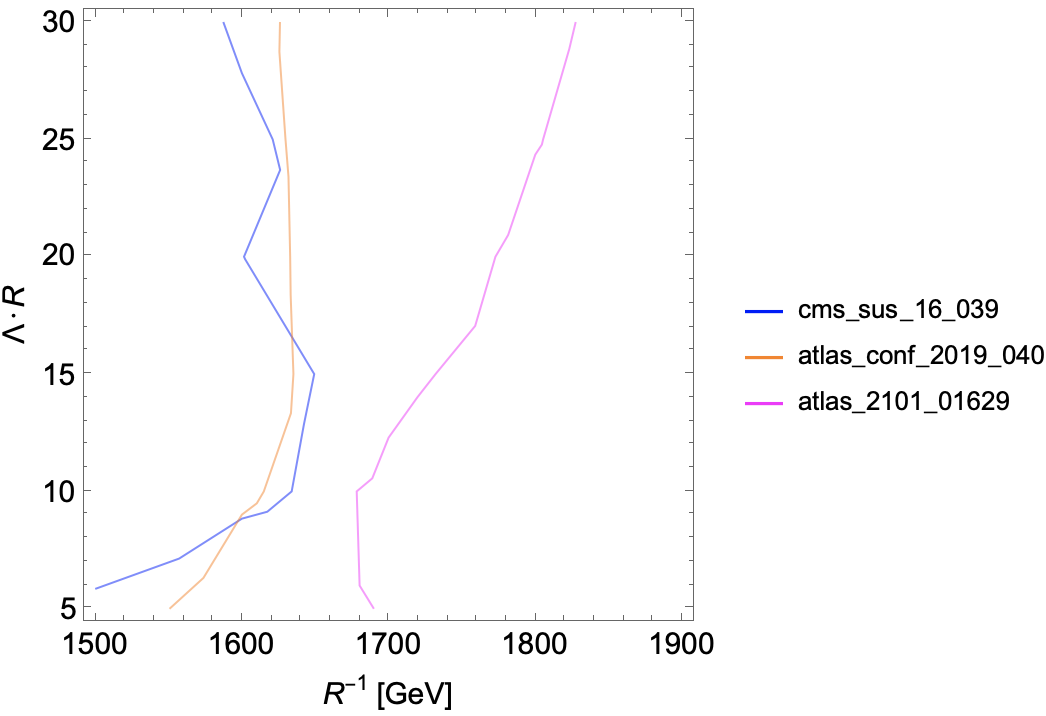}
\caption{Exclusion plot for the searches listed in Table~\ref{tab:analysis} obtained with the MLM matching and the default shower setting, as discussed in Section~\ref{S:4}. The excluded region is to the left of the lines.}
\label{fig:analyses}
\end{figure}

\begin{figure}[t]
\centering\includegraphics[width=1.0\linewidth]{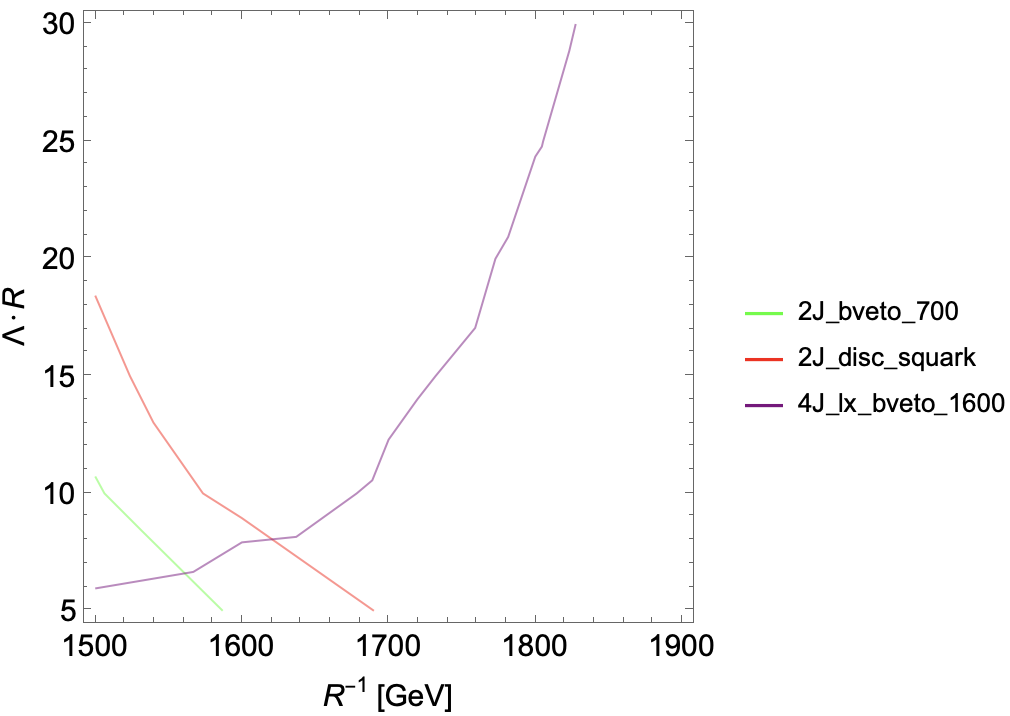}
\caption{The most sensitive signal regions comprising the \texttt{atlas\_2101\_01629} search. The corresponding labels are the ones used by \textsc{CheckMATE}, see text for details. The excluded region is to the left of the lines.}
\label{fig:signalregions}
\end{figure}

Figure~\ref{fig:analyses} compares limits from the searches listed in Table~\ref{tab:analysis}, which have the highest sensitivity to the MUED signal. We see that the strongest limit comes from the \texttt{atlas\_2101\_01629} search for any value of $\Lambda\cdot R$. Figure~\ref{fig:signalregions} shows the three best signal regions that provide the strongest exclusion for this particular search. For $\Lambda\cdot R < 10$, the most sensitive signal region (SR) is 2J squark discovery SR (\textsc{CheckMATE} label \texttt{2J\_disc\_squark}) which requires at least two jets and a low-$p_T$ lepton: $p_T > 7 (6)$ GeV for electron (muon) and $p_T < 20$ or $25$ GeV, depending on a number of jets. The requirements placed on missing transverse momentum and effective mass, $E_\mathrm{T}^\mathrm{miss} > 400$~GeV, $m_\mathrm{eff} > 1200$~GeV, enhance the sensitivity by selecting signal events with boosted final-state particles recoiling against energetic ISR jets. It is therefore well-suited for probing parameter space with low mass splitting, as it is the case for $\Lambda\cdot R < 10$. As we go up in $\Lambda\cdot R$, the sensitivity then shifts to the 4J low-x SR (\textsc{CheckMATE} label \texttt{4J\_lx\_bveto\_1600}) which corresponds to a higher jet multiplicity (with no $b$-jets) of 4 to 5, and effective mass $m_\mathrm{eff} > 1600$~GeV, which is expected in the larger mass splitting region of higher $\Lambda\cdot R$.

It is also illuminating to look at the next two most sensitive searches just to understand the shape of exclusion lines. For the search \texttt{cms\_sus\_16\_039}, the most sensitive signal regions is \texttt{SR\_A14} (\texttt{SR\_A09}) where events are required to have three electrons or muons that form at least one opposite-sign same-flavor pair and with $E_\mathrm{T}^\mathrm{miss}$ $\geq$ 250 GeV (200 GeV). Notably, the multi-lepton search using $35.9$ fb$^{-1}$ of data reaches similar sensitivity to the multi-jet search \texttt{atlas\_conf\_2019\_040} using $139$ fb$^{-1}$. In case of the latter search, the most sensitive signal region is \texttt{SR\_4j\_1000} which requires at least 4 jets, the leading jet $p_T > 200$ GeV, $m_\mathrm{eff} > 1000$ GeV and $E_\mathrm{T}^\mathrm{miss}/\sqrt{H_\mathrm{T}} > 16$.\footnote{We note these limits are different from those obtained in Ref.~\cite{indianpaper3}.}

As already suggested, the leptonic searches can be expected to have an improved sensitivity to MUED due to the relatively large expected number of leptons in the final state compared to SUSY. Indeed, our study finds the 1-lepton search, \texttt{atlas\_2101\_01629}, having the best sensitivity in the full data set but the CMS multilepton search, \texttt{cms\_sus\_16\_039} using $\mathcal{L} = 36\ \mathrm{fb}^{-1}$, was also very promising and it is likely that a similar study using the full data set will be at least as sensitive as the ATLAS search. We note that in Ref.~\cite{ATLAS:2015gky} it was found that same-sign/3-lepton search~\cite{ATLAS:2014kpx} was the most constraining for moderate to large values of $\Lambda\cdot R$. In our study, however, the ATLAS same-sign leptons study, \texttt{atlas\_1909\_08457}, was lagging behind in sensitivity. Similarly, at low $\Lambda\cdot R$ Ref.~\cite{ATLAS:2015gky} points to a  search based on 2 soft leptons~\cite{ATLAS:2015rul}. Again, it turns out that the full luminosity search for soft-leptons final states, \texttt{atlas\_1911\_12606}~\cite{ATLAS:2019lng}, is not as strong as the 1-lepton search \texttt{atlas\_2101\_01629}.  

\begin{figure}[h]
\centering\includegraphics[width=1.0\linewidth]{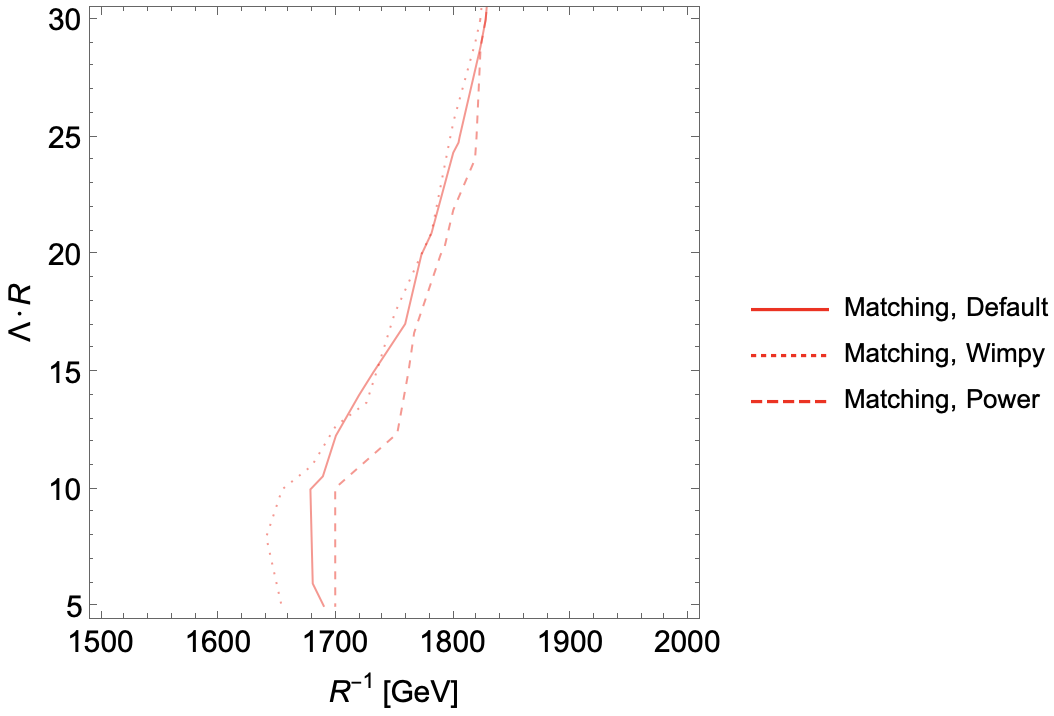}
\caption{The combined exclusion in the minimal UED model obtained with the matched samples and different shower settings discussed in the text. Note that the exclusion is entirely due to the \texttt{atlas\_2101\_01629} search. The excluded region is to the left of the line.}
\label{fig:asymptote}
\end{figure}

Finally, in Fig.~\ref{fig:asymptote} we summarize our findings. It shows the exclusion line of the MUED model obtained with the matched samples and different shower settings. The exclusion in $1/R$ shows little variation at large $\Lambda\cdot R$, stabilizing around 1860 GeV. Note that most NLO calculations are missing in the literature except for KK gluons\cite{Freitas:2017hov}. The correction is modest, up to 14\%. Uncertainty due to cross section was estimated in Ref.~\cite{Freitas:2017hov} to be $^{+5\%}_{-8\%}$ at the NLO and $^{+15\%}_{-17\%}$ and was accounted for in \textsc{CheckMATE}. 

In the LHC-allowed region above $R^{-1} \sim 1700$ GeV the dark matter relic density is too high~\cite{Cornell:2014jza}. Thus, the LHC constraints together with the dark matter relic density bound essentially rule out MUED, at least for the standard cosmology.

\section{Conclusion}
\label{S:5}

We showed that in updating the limits for the MUED model, using parton showers alone yields large uncertainty at lower $\Lambda\cdot R$, i.e.\ for compressed mass spectra. This clearly demonstrates the case for employing an appropriate matching procedure. We then updated existing limits using the current LHC searches concluding that the strongest limit comes from the \texttt{atlas\_2101\_01629} search, which targets the final states with one lepton, jets and missing transverse energy.



\section*{Appendix A: List of \textsc{CheckMATE} Analyses}

Table~\ref{tab:analyses} shows the list of 13 TeV ATLAS and CMS analyses implemented into \textsc{CheckMATE} as of completing this study. 

\begin{table}[h!]
\caption{A list of $\sqrt{s} = 13$ TeV analyses implemented in the  \textsc{CheckMATE} version used throughout this work. The last column shows the total integrated luminosity.}
\centering
\begin{tabular}{|l|l|l|}
\hline
Analysis               & $\mathcal{L}$ {[}fb$^{-1}${]} \\ \hline
\texttt{atlas\_1604\_01306} \cite{ATLAS:2016zxj} & 3.2 \\
\texttt{atlas\_1605\_09318} \cite{ATLAS:2016gty} & 3.2 \\
\texttt{atlas\_1609\_01599 } \cite{ATLAS:2016wgc} & 3.2 \\
\texttt{atlas\_1704\_03848} \cite{ATLAS:2017nga} & 36.1 \\
\texttt{atlas\_1706\_03731} \cite{ATLAS:2017tmw} & 36.1 \\
\texttt{atlas\_1708\_07875} \cite{ATLAS:2017qwn} & 36.1 \\
\texttt{atlas\_1709\_04183} \cite{ATLAS:2017drc} & 36.1 \\
\texttt{atlas\_1712\_02332} \cite{ATLAS:2017mjy} & 36.1 \\
\texttt{atlas\_1712\_08119} \cite{ATLAS:2017vat} & 36.1 \\
\texttt{atlas\_1802\_03158} \cite{ATLAS:2018nud} & 36.1 \\
\texttt{atlas\_1803\_02762} \cite{ATLAS:2018ojr} & 36.1 \\
\texttt{atlas\_1908\_08215} \cite{ATLAS:2019lff} & 139 \\
\texttt{atlas\_1909\_08457} \cite{ATLAS:2019fag} & 139 \\
\texttt{atlas\_1911\_12606} \cite{ATLAS:2019lng} & 139 \\
\texttt{atlas\_2101\_01629} \cite{atlas2101}      & 139 \\
\texttt{atlas\_conf\_2015\_082} \cite{ATLAS-CONF-2015-082} & 3.2 \\
\texttt{atlas\_conf\_2016\_013} \cite{ATLAS-CONF-2016-013} & 3.2 \\
\texttt{atlas\_conf\_2016\_050} \cite{ATLAS-CONF-2016-050} & 13.2 \\
\texttt{atlas\_conf\_2016\_054} \cite{ATLAS-CONF-2016-054} & 14.8 \\
\texttt{atlas\_conf\_2016\_066} \cite{ATLAS-CONF-2016-066} & 13.3 \\
\texttt{atlas\_conf\_2016\_076} \cite{ATLAS-CONF-2016-076} & 13.3 \\
\texttt{atlas\_conf\_2016\_096} \cite{ATLAS-CONF-2016-096} & 13.3 \\
\texttt{atlas\_conf\_2017\_060} \cite{ATLAS-CONF-2017-060} & 36.1 \\
\texttt{atlas\_conf\_2019\_020} \cite{ATLAS-CONF-2019-020} & 139 \\
\texttt{atlas\_conf\_2019\_040} \cite{atlas040} & 139 \\
\texttt{cms\_pas\_sus\_15\_011} \cite{CMS-PAS-SUS-15-011} & 2.2 \\
\texttt{cms\_sus\_16\_025} \cite{CMS-PAS-SUS-16-025} & 12.9 \\
\texttt{cms\_sus\_16\_039} \cite{cms039}      & 35.9  \\ 
\texttt{cms\_sus\_16\_048} \cite{CMS-PAS-SUS-16-048} & 35.9 \\
\hline
\end{tabular}
\label{tab:analyses}
\end{table}

\section*{Acknowledgements}
MF is funded by the UP System Enhanced Creative Work and Research Grant (ECWRG-2020-2-8R). KR is supported by the National Science Centre, Poland under grants: 2016/23/G/ST2/04301, 2015/18/M/ST2/00518, 2018/31/B/ST2/02283, 2019/35/B/ST2/02008. R. RdA acknowledges partial funding/support from the Elusives ITN (Marie Sk\l{}odowska-Curie grant agreement No 674896) and the ``SOM Sabor y origen de la Materia" (FPA 2017-85985-P).

\bibliographystyle{JHEP}
\bibliography{main}

\end{document}